\def\ln{\ell{n}}
\documentstyle[12pt]{article}

 {\parindent 0.5cm \textwidth 15.0cm \textheight
23.5cm \hoffset=-0.8cm \voffset=-3cm
  \let\LARGE=\large
 \let\large=\normalsize
\newcommand{\be}{\begin{equation}}
\newcommand{\ee}{\end{equation}}
\newcommand{\ba}{\begin{array}{c}}
\newcommand{\ea}{\end{array}}

\begin{document}
\begin{titlepage} \vspace{0.2in} \begin{flushright}
MITH-95/1 \\ \end{flushright} \vspace*{1.5cm}
\begin{center} {\LARGE \bf  The Standard Model on a Planck Lattice:\\
Mixing Angles vs Quark Masses\\} \vspace*{0.8cm}
{\bf Giuliano Preparata$^{(a,b)}$ and She-Sheng Xue$^{(a)}$}\\ \vspace*{1cm}
(a) INFN - Section of Milan, Via Celoria 16, Milan, Italy\\
(b) Dipartimento di Fisica, Universit\'a di Milano, Italy\\ \vspace*{1.8cm}
{\bf   Abstract  \\ } \end{center} \indent
Proceeding in our research programme on the Standard Model (SM) on a Planck
lattice (PLSM) we analyse the emergence of the quark masses through a
``superconducting'' mechanism of spontaneous violation of the chiral symmetry
of the Standard Model. Due to the peculiar structure of the lattice coupling of
the $W^\pm$ gauge bosons, we discover four new relationships between the angles
of the mixing matrix and the (current) quark masses. In particular, we predict
the CP-violating phase $\delta_{13}\simeq 79^\circ$.

\vfill \begin{flushleft}  January, 1995   \vspace*{3cm} \\
\noindent{\rule[-.3cm]{5cm}{.02cm}} \\
\vspace*{0.2cm} \hspace*{0.5cm} ${}^{a)}$
E-mail address: xue@milano.infn.it\end{flushleft} \end{titlepage}

\noindent
{\bf 1.}\hspace*{0.3cm} The Standard Model on a Planck lattice (PLSM) is a
research programme that we are developing out since a few years with the aim of
finding out whether some of the unsatisfactory aspects of the continuum
Standard Model (CSM) might be overcome in the new formulation. A formulation,
it should be strongly emphasized, that does not stem from some more or less
capricious conjecture, but is motivated by the definite expectation, first
spelled out by J.A.~Wheeler\cite{wheeler}, that the violent quantum
fluctuations of the gravitational field at the Planck scale may well ``tear
apart'' the continuous space-time of the CSM to yield a discrete structure,
which may be conveniently modeled by a regular four-dimensional lattice, of
lattice constant $a=a_p \simeq 10^{-33}cm$.

The remarkable aspect of the PLSM is the necessity it engenders to go beyond
the simple lattice transcription of the CSM, as demanded by the well known
``no-go theorem'' of Nielsen and Ninomiya\cite{nogo}. And, as we have discussed
at length elsewhere\cite{x1}, in order to avoid the difficulties of the ``no-go
theorem'' one needs simply add to the lattice transcription of the action of
the CSM - bilinear in the Fermi fields - two well defined terms, quadrilinear
in the Fermi fields\footnote{It should be stressed that terms of this type are
indeed expected in a theory that includes gravitation, for they represent
effective gravitational interactions among matter-fields at the Planck scale.},
akin to those first introduced in the context of spontaneous chiral symmetry
breaking by Nambu and Jona-Lasinio (NJL)\cite{nambu}. As a result of such
addition, not only have we been able to evade the inconsistencies of the
``no-go theorem'', but we have obtained explicit mechanisms to violate
(spontaneously) the basic chiral invariance of the SM action, without the
introduction of totally extrinsic new forms of matter, such as the Higgs
doublets of the CSM.

A careful analysis of the structure of the vacuum has shown\cite{x2} that in
terms of the NJL-terms {\it only} (i.e.~neglecting at a first stage of the
analysis the interactions of the Fermi-fields with the SM gauge fields) there
exists a ground state (depending on ``fine-tuned'' values of the adimensional
coupling constants $g_1$ and $g_2$, associated with the two NJL-terms) in which
all fermions (quarks and leptons) except a single quark generation (to be
called t- and b- quarks) remain massless. The four massless bosons that,
according to Goldstone theorem, are associated with such spontaneously broken
vacuum, as well known, completely disappear when the gauge-interactions are
turned on, producing the longitudinal polarization states of the $W^\pm$ and
$Z^\circ$ bosons, that thus acquire masses, which are related in a well
defined way to the masses of the t- and b-quarks \cite{bar}.

As shown in ref.\cite{x4}, turning the gauge interactions on and assuming that
the weak interaction doublets exactly coincide with the quark generation
doublets, i.e.~that {\it there is no weak mixing} among the quark generations,
one obtains a quarks mass pattern (that we shall call the ``ideal'' pattern)
characterized by ${m_t\over m_b}\sim 30$ and by the
masslessness of the rest of the quarks: $m_u=m_d=m_s=m_c=0$.
A remarkable aspect of the ``ideal'' (unmixed) pattern of the quark masses is
clearly its closeness to the experimental one, in which the
would-be massless quarks happen to be all confined in a narrow mass range
between a few MeV (u,d quarks) and somewhere around 1 GeV (the c-quark).

In this letter we will show that in the PLSM the non-trivial observed masses of
the would-be massless quarks can be explained by the real world being, unlike
the ``ideal'' one, mixed and that the small deviations of the real from the
ideal pattern fully determine the rotation angles
$(\theta_{12},\theta_{23},\theta_{13})$\footnote{ In the
Cabibbo-Kobayashi-Maskawa (C-K-M)
representation of the mixing matrix\cite{km}}, that, as experimentally
observed, turn out to be quite small and hierarchically arranged
$(\theta_{12}>>\theta_{23}>>\theta_{13})$. Finally an approximate prediction of
the CP-violating phase will be given in terms of the quark masses.

\vspace*{0.5cm}
\noindent
{\bf 2.}\hspace*{0.3cm} In \cite{x4}, we have shown that the Dyson equations
for quark
mass matrices $\Sigma (p)$ in the low-energy region $(p<<\Lambda)$ can be
represented by the following diagram,
\vspace*{4cm}
\begin{equation}
\label{dia}
\end{equation}
where the $O({1\over a})$ contributions to $\Sigma(p)$ are absent due to the
Ward identies stemming from the basic chiral symmetry of the PLSM
\cite{rome}. In the R.H.S.~of the above equation (\ref{dia}), the
first two terms are contributions from NJL interactions and the second term
arises from the gauge-boson (except $W^\pm$) contributions, the last term,
which is absent in the CSM, is a genuine lattice $W^\pm$-contribution to the
quark mass matrices. This $W^\pm$-contribution\cite{x4} stems from the
non-chiral gauge coupling of $W^\pm$ to fermions $\Gamma_\mu=\gamma_\mu P_L\cos
(p_\mu a)+ r\sin(p_\mu a)$, where $r$ is the chiral-symmetry-violating Wilson
parameter\cite{wilson}, $P_L={1\over2}(1-\gamma_5)$ and $a$ is the Planck
length.

As a first step of approximation, one can take the large-$N_c$ and
mean-field approximation to the Dyson equations (\ref{dia}) for the quark
masses
$\Sigma(p)\simeq m$, obtaining
\begin{eqnarray}
Q^f_{2\over3}m^f_{2\over3}=\Sigma_l|V_{fl}|^2m^l_{-{1\over3}},\hskip1cm
f=u,c,t\nonumber\\
Q^l_{-{1\over3}}m^l_{-{1\over3}}=\Sigma_f|V_{lf}|^2m^f_{2\over3},\hskip1cm
l=d,s,b,
\label{gap}
\end{eqnarray}
where $m^f_{2\over3}=m_u,m_c,m_t$ and $m^l_{-{1\over3}}=m_d,m_s,m_b$: the quark
masses for the $e={2\over3}$ sector and $e={-{1\over3}}$ sector respectively.
The R.H.S.~of eq.(\ref{gap}) syems from the $W^\pm$-exchange term in
eq.(\ref{dia}) and $V_{fl}$ are elements of the C-K-M matrix. Due to the
non-tivial nature of the mixing matrix, the six Dyson equations (\ref{gap})
become coupled equations. The factors $Q^f_{2\over3}$ and $Q^l_{-{1\over3}}$ in
(\ref{gap}) are in general quite complicated, for instance one has
\begin{equation}
Q^f_{2\over3}=\bar L_w^{-1}(r)[1-NJL(g_{1,2},\Lambda,\ln{\Lambda\over
m^f_{2\over3}})
-gauge(g,\ln{\Lambda\over m_f})],
\label{q}
\end{equation}
where the function $\bar L_w(r)$ is given in the ref.\cite{x4}, {\it ``NJL''}
represents the contributions of NJL interactions and {\it ``gauge''} stands for
the contribution of the $\gamma-Z^\circ$ gauge interactions. As we have
stressed in the Introduction, if we omitted the gauge contibution, we would
obtain the solution $m_u=m_d=m_s=m_c=0$, and $m_t=m_b\not=0$, for the minimum
of the ground state energy occurs when only one quark generation becomes
massive, and the symmetry of the NJL-terms requires the equality of the top and
bottom masses. When we include in (\ref{gap}) the gauge-interactions with the
without mixing, we get the solution of the ``ideal'' pattern,
\begin{eqnarray}
 m_t&\sim& 30m_b\not=0\\
m_u&=&m_c=m_d=m_s=0,
\label{tb}
\end{eqnarray}
where $m_t\not=m_b$ is due to the fact that top and bottom quarks have
different electric charges.

In this article we wish to go further and see whether the ``ideal'' pattern may
become the ``real'' pattern, due to the small (observed) deviation of the C-K-M
matrix from triviality. We should emphasise at this point that in the Standard
Model (either the PLSM or the CSM) the C-K-M matrix is a totally extrinsic
element, qualifying our {\it actual} world as compared (and opposed) to any
other {\it possible} world, where the C-K-M can be either trivial (the
``ideal'' world) or anything one wishes. We consider it already a definite
(qualitative) success of the PLSM to have predicted an ``ideal'' pattern so
close to the ``real'' one, implying in turn that we do live in a world where
the C-K-M matrix is {\it almost} trivial.

Since the factors $Q^f_{2\over3}$ and $Q^l_{-{1\over3}}$ only weakly
(logrithmically) depend on quark masses, in eqs.(\ref
{gap}) we make the approximations,
\begin{equation}
Q^u_{2\over3}\simeq Q^c_{2\over3}\simeq Q^t_{2\over3};\hskip1cm
Q^d_{-{1\over3}}\simeq Q^s_{-{1\over3}}\simeq Q^b_{-{1\over3}}.
\label{qq}
\end{equation}
Thus, from the six Dyson
equations (\ref{gap}), we obtain four independent equations
\begin{eqnarray}
{m_u\over m_c}&=&{|V_{ud}|^2m_d+|V_{us}|^2m_s+|V_{ub}|^2m_b\over
|V_{cd}|^2m_d+|V_{cs}|^2m_s+|V_{cb}|^2m_b},\label{r1}\\
{m_u\over m_t}&=&{|V_{ud}|^2m_d+|V_{us}|^2m_s+|V_{ub}|^2m_b\over
|V_{td}|^2m_d+|V_{ts}|^2m_s+|V_{tb}|^2m_b},\label{r2}\\
{m_d\over m_s}&=&{|V_{du}|^2m_u+|V_{dc}|^2m_c+|V_{dt}|^2m_t\over
|V_{su}|^2m_u+|V_{sc}|^2m_c+|V_{st}|^2m_t},\label{r3}\\
{m_d\over m_b}&=&{|V_{du}|^2m_u+|V_{dc}|^2m_c+|V_{dt}|^2m_t\over
|V_{bu}|^2m_u+|V_{bc}|^2m_c+|V_{bt}|^2m_t}.\label{r4}
\end{eqnarray}
These four equations determine completely the mixing matrix in terms of the
quark masses. Using the standard parameterisation of the C-K-M matrix
\cite{pkm}
with four mixing angles $\theta_{12}(\theta_c),\theta_{23}, \theta_{13}$ and
$\delta_{13}$ (the CP-violating phase), one can obtain definite, though rather
complicated
relationships between such mixing angles and the pattern of quark masses.
Defining, for convenience
\begin{eqnarray}
x&=&\tan^2\theta_c;\hskip1cm y=\sin^2\theta_{23}\nonumber\\
z&=&\cot^2\theta_{13};\hskip1cm w=\cos\delta_{13},\nonumber
\end{eqnarray}
one gets,
\begin{eqnarray}
{m_u\over m_c}&=&{z(m_d+xm_s)+(1+x)m_b\over (xm_d{+}m_s)(1{-}y)(1{+}z)
{+}y(m_d{+}xm_s){+}(1{+}x)yzm_b{+}(m_d{-}m_s)wc},\label{cos}\\
{m_u\over m_t}&=&{z(m_d+xm_s)+(1+x)m_b\over y(1{+}z)(xm_d{+}m_s){+}(1{+}y)
[m_d{+}xm_s
{+}(1{+}x)zm_b]{+}(m_s{-}m_d)wc},\label{tan}\\
{m_d\over m_s}&=&{zm_u{+}[(1{+}z)x(1{-}y){+}y]m_c{+}[xy(1{+}z){+}1{-}y]m_t{+}
(m_c{-}m_t)wc\over
xzm_u{+}[(1{-}y)(1{+}z){+}xy]m_c{+}[y(1{+}z){+}x(1{-}y)]m_t{+}(m_t{-}m_c)wc},\label{sin}\\
{m_d\over m_b}&=&{zm_u{+}[(1{+}z)x(1{-}y){+}y]m_c{+}[xy(1{+}z){+}1{-}y]m_t{+}
(m_c{-}m_t)wc\over
(1+y)[m_u+z(ym_c+(1-y)m_t)]},
\label{cot}
\end{eqnarray}
where
\begin{equation}
c=2\sqrt{xy(1-y)(1+z)}.
\label{c}
\end{equation}
These awful looking equations can be drastically simplified if one takes into
account the observed hierarchical pattern of quark masses, i.e.~the fact that
the third
generation (t,b) is much heavier than the others. One can then approximately
solve the above four independent equations (\ref{tan})-(\ref{cot}) and get
\begin{eqnarray}
tan^2\theta_c&\simeq& -{m_d\over m_s};\hskip1cm\sin^2\theta_{23}\simeq
-{m_c\over m_t},\label{res1}\\
\cot^2\theta_{13}&\simeq& {m_t\over m_u};\hskip1cm \cos\delta_{13}\simeq
{1\over 2}\sqrt{m_sm_u\over m_cm_d},\label{res2}
\end{eqnarray}
where the strange and ominous looking minus signs, can all be eliminated by
chirally rotating ( an operation that the symmetry of the PLSM allows) the
fields of the (c, s) generation so as to transform $-m_s, -m_c\rightarrow m_s,
m_c$. We note that $\tan\theta_c\simeq{m_d\over m_s}$ in eq.(\ref{res1}) has
been around in completely different contexts for a quarter of a century\cite{f}
now, while the others - to our knowledge - appear to be genuine consequences of
the peculiar chiral symmetry breaking of the PLSM. The most interesting aspect
of eqs.(\ref{res1}) and (\ref{res2}) is their tying the observed hierarchical
pattern of the inter-generation mixing angles
$\theta_c>\theta_{23}>\theta_{13}$, to the observed hierarchy of mass ratios
${m_s\over m_d}>>{m_c\over 2m_t}>>{m_u\over m_t}$. Finally, as for the
CP-violating phase $\delta_{13}$, eq.(\ref{cos}) can be cast in the form
\begin{equation}
\cos\delta_{13}\simeq{1\over2}{\sin\theta_{13}\cos\theta_{23}\cos\theta_c\over
\sin\theta_{23}\sin\theta_c},
\label{cp}
\end{equation}
which by using the
experimental determinations of the mixing angles (central values
$\sin\theta_c=0.221; \sin\theta_{23}=0.040; \sin\theta_{13}=0.0035$\cite{data})
yields the prediction
\begin{equation}
\delta_{13}\simeq 79^\circ.
\label{cpa}
\end{equation}

\vspace*{0.5cm}
\noindent
{\bf 3.}\hspace*{0.3cm}
To conclude, we have shown that within a novel approach to the physics of the
SM - the PLSM - it is possible to give a full description of the quark mass
spectrum in terms of a well defined ``superconductive'' mechanism (the NJL
four-fermion interactions), the standard gauge interactions ($\gamma, W^\pm,
Z^\circ$) and the mixing matrix, which in our universe defines the alignment
between the generation doublets and the pure weak isospin doublets. In this way
we have derived a set of relationships between the (current) quark masses and
the four angles that parameterise the mixing matrix which agree with
observation. Of particular interest is the prediction (\ref{cpa}) of the
CP-violating phase.

}
\end{document}